\documentclass[twocolumn,preprintnumbers,amsmath,amssymb,aps]{revtex4}

\usepackage{graphicx}
\usepackage{dcolumn}
\usepackage{bm}

\begin{document}

\title{Bistability and oscillatory motion of natural nano-membranes appearing within monolayer graphene on silicon dioxide}

\author{T. Mashoff$^1$}
\author{M. Pratzer$^1$}
\email[corresponding author: ]{pratzer@physik.rwth-aachen.de}
\author{V. Geringer$^1$}
\author{T. J. Echtermeyer$^2$}
\author{M. C. Lemme$^2$}
\author{M. Liebmann$^1$}
\author{M. Morgenstern$^1$}
\affiliation{$^1$ 2nd Institute of Physics B , RWTH Aachen University, 52074 Aachen, Germany; J\"ulich-Aachen Research Alliance: Fundamentals of Future Information Technology (JARA-FIT)\\
$^2$ Advanced Microelectronic Center Aachen (AMICA), AMO GmbH, Otto-Blumenthal-Stra{\ss}e 25, 52074 Aachen, Germany}

\date{\today}

\begin{abstract}

The recently found material graphene is a truly two-dimensional crystal and exhibits, in addition, an extreme mechanical strength. This in 
combination with the high electron mobility favours graphene for electromechanical investigations down to the quantum limit. Here, we show that a 
monolayer of graphene on SiO${_2}$ provides natural, ultra-small membranes of diameters down to 3\,nm, which are caused by the intrinsic rippling 
of the material. Some of these nano-membranes can be switched hysteretically between two vertical positions using the electric field of the tip 
of a scanning tunnelling microscope (STM). They can also be forced to oscillatory motion by a low frequency ac-field. Using the mechanical 
constants determined previously \cite{Lee}, we estimate a high resonance frequency up to 0.4\,THz. This might be favorable for 
quantum-electromechanics and is prospective for single atom mass spectrometers.

\end{abstract}

\maketitle

Nano-electromechanical systems (NEMS) are promising, e.g., as ultra-low mass detectors. One measures the shift in resonance frequency of a micro- 
or nano-object, when a particle adsorbs onto it. Mostly, silicon nano-beams with sensitivities down to attograms are used \cite{Bunch, Lavrik, 
Ilic, Peng}. For further improvement of sensitivity, the oscillators have to become lighter, e.g., thinner. Graphene is one atomic layer thick 
and exhibits an extreme mechanical strength \cite{Bunch, Peng, Jensen, Bunch2} which could be ideal. Resonators using double layer graphene have 
already been demonstrated implying a mass sensitivity of $4.5\times 10^{-22}$\,kg \cite{Bunch}. The highest sensitivity so far is found for a 
double wall carbon nanotube, which exhibits resonance frequency shifts compatible with the mass of one gold atom \cite{Jensen}.

In this work, we study a graphene monolayer on SiO$_2$ \cite{Novo} by STM. We find movable areas within the valleys of the intrinsic rippling 
\cite{Geringer,Meyer} exhibiting an extremely small size of $6-21$\,nm$^2$ ($230-800$ carbon atoms). They show either a hysteretic or a 
reversible  deflection in response to the tip bias or a change in tip-graphene distance. The areas can be forced to oscillatory motion by an 
ac-field. Bistability and oscillatory motion are reproduced using the clamped membrane model \cite{Lee,Komaragiri,Wan} including electrostatic 
and van-der-Waals (vdW) forces of tip and substrate. We deduce resonance frequencies up to 400\,GHz corresponding to a vibronic energy of 1.6 
meV. Such a large value might provide easy cooling to the ground state leading to a novel access to quantum-mechanical manipulation of vibrons 
\cite{Kleckner,Anghel} or, in combination with the low membrane mass, to mass detection of single hydrogen atoms.

\begin{figure*}
\includegraphics[width=18cm]{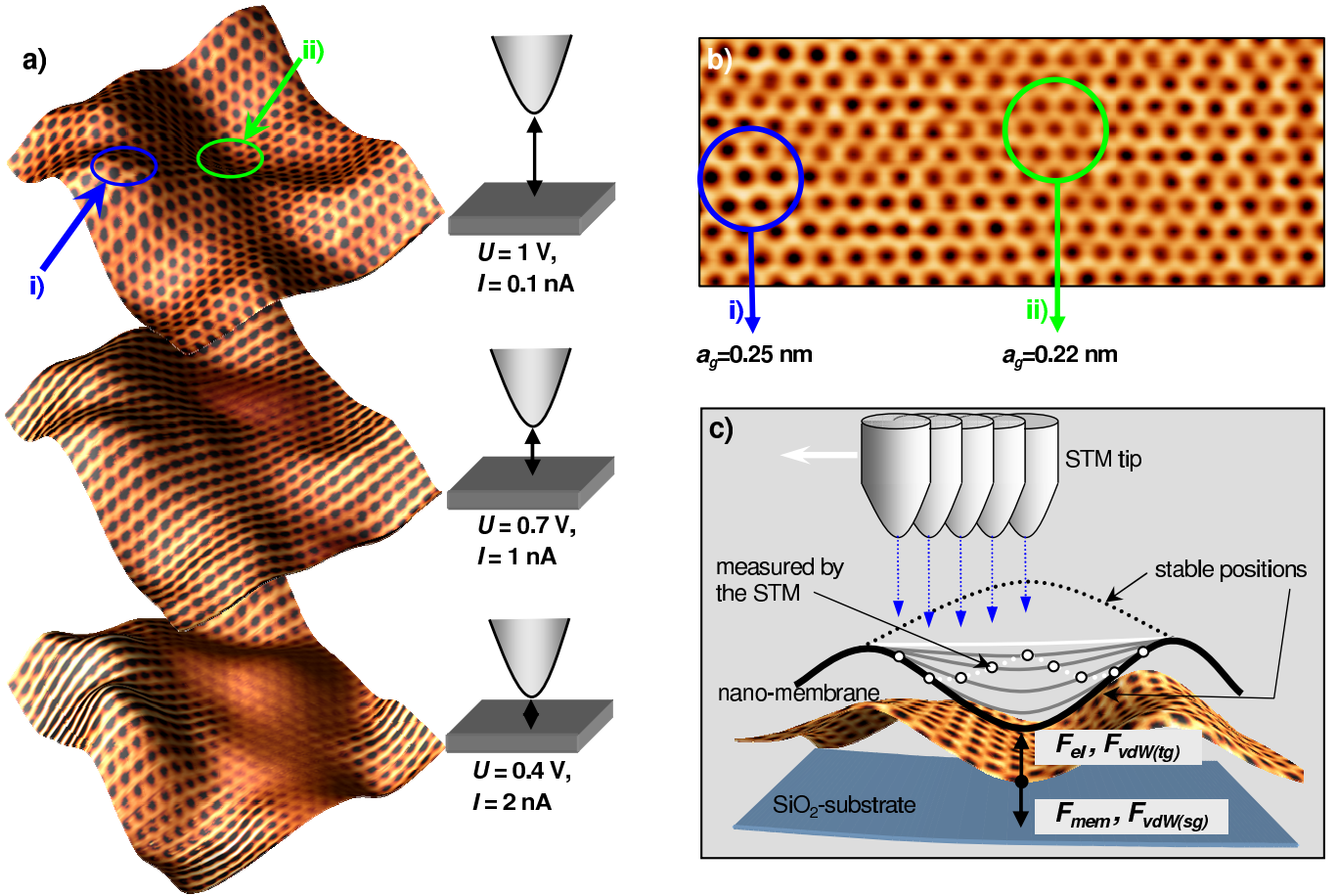}
\caption{\label{lifting}Reversible lifting of the graphene nano-membrane. a) 3D representations of an atomically resolved monolayer graphene at 
different tip-sample distance as indicated on the right. At a large distance ($U=1$\,V, $I=0.1$\,nA) carbon hexagons are visible. Decreasing the 
distance, a hill appears in the centre of the valley and the hexagons are transformed to bumps at every second atom position. b) High-pass 
filtered STM image ($U=1$\,V and $I=0.1$\,nA). Circles denote the hill and valley areas marked in a. The apparent lattice constants show a 
difference of $\Delta a=0.03$\,nm due to the tilting of the $\pi$-orbitals. c) Model explaining the mechanical behaviour of the nano-membrane. Due to 
the locally applied force between tip and graphene, the valley is continuously lifted while scanning the tip and the STM image shows a hill in 
the centre of the valley (dotted line).}
\end{figure*}
\begin{figure*}
\includegraphics[width=18cm]{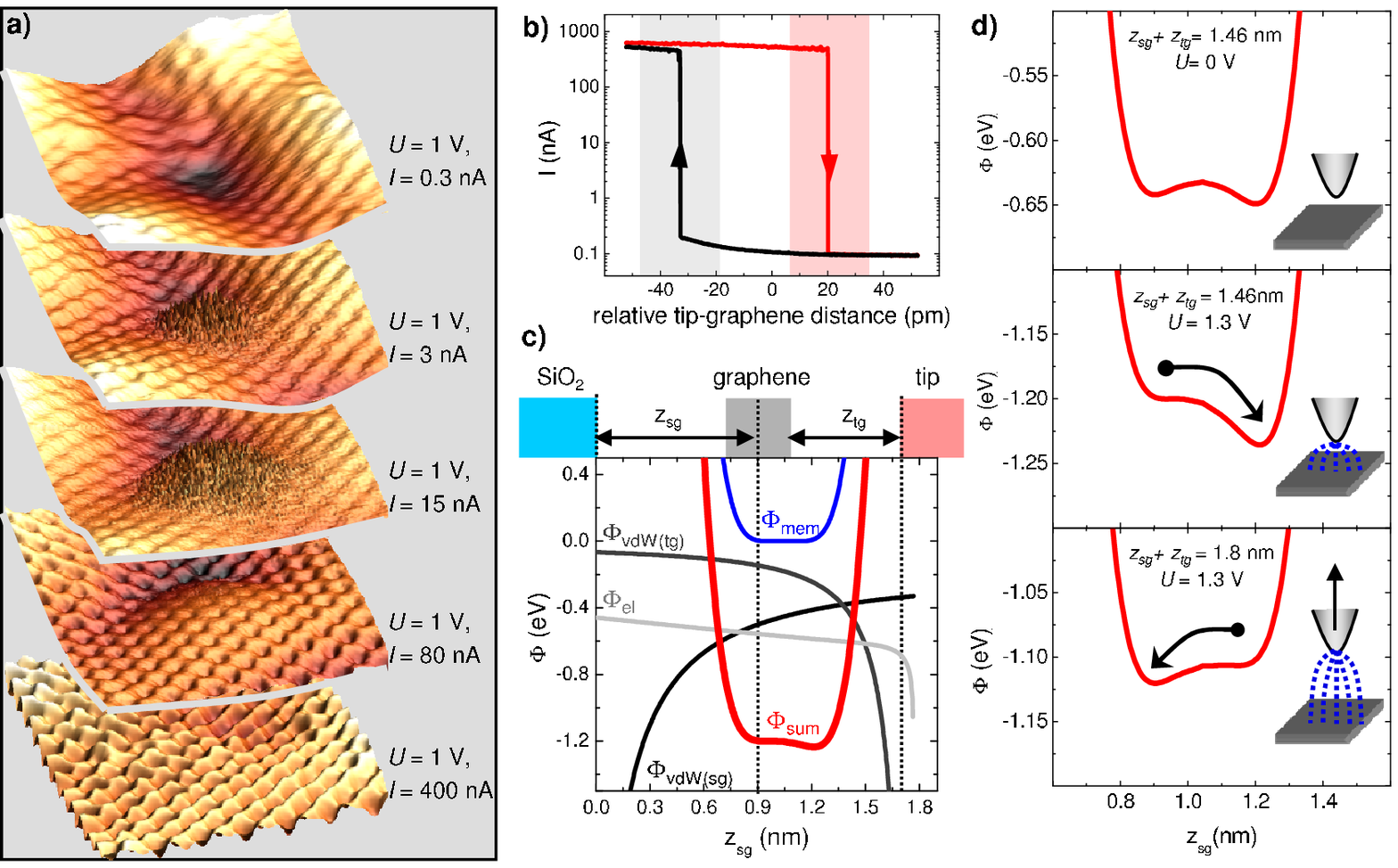}
\caption{\label{hysteresis}Hysteretic nano-membrane. a) Constant-current images of a graphene area taken at different tip-surface distances. At 
intermediate distance ($I=3$\,nA, $15$\,nA), a part of the valley exhibits unstable behaviour visible as spikes. Higher current leads to stable 
imaging again, but the valley has turned into a hill. b) I(z)-spectrum revealing  hysteretic behaviour. c) Calculated potential energy $\Phi_{\rm 
sum}$ as a function of substrate-graphene distance z$_{\rm sg}$ at fixed tip-substrate distance. $\Phi_{\rm 
sum}$ consists of electrostatic potential $\Phi_{\rm el}$, van-der-Waals potentials between tip and graphene $\Phi_{\rm vdW(tg)}$ and SiO$_2$ 
and graphene $\Phi_{\rm vdW(sg)}$, and the elastic membrane potential $\Phi_{\rm mem}$. The two local minima in $\Phi_{\rm sum}$ indicate two 
metastable positions. The definition of distances is sketched at the top. d) Using an initial tip-substrate distance $z_{\rm tg}+z_{\rm sg}=1.46$\,nm, (top) the hysteretic switching in constant-current mode can be reproduced by first applying a voltage moving the membrane towards the tip (middle) and then retracting the tip to reduce current again, thereby switching the membrane back (bottom).}
\end{figure*}
The morphology of graphene has been studied previously by STM \cite{Geringer, Ishigami, Stolyarova, Zhang}. By comparing with the morphology of 
SiO${_2}$, we revealed that graphene can exhibit an intrinsic rippling not induced by the substrate. We argued that the rippling appears, if 
graphene is freely suspended between hills of the SiO$_2$ \cite{Geringer}. The rippling (amplitude: 0.3-0.5\,nm) exhibits a preferential wave 
length of 15\,nm at 300\,K, \cite{Geringer} slightly increased at 4.8\,K to 20\,nm. At smaller length scales, we observe additional corrugation 
of $\pm$70\,pm. Figure \ref{lifting}a shows this corrugation at several tip-sample distances. At large distance (top), the corrugated surface and 
the typical carbon hexagons are visible. Interestingly, the apparent lattice constant differs by 14\,\% between oppositely curved areas (Fig. 
\ref{lifting}b). This is explained by the tilting of the p$_z$-orbitals within the curved surface assuming an effective p$_z$-length of $0.26\pm 
0.05$\,nm in accordance with theory \cite{Balatsky}.

Decreasing the tip-graphene distance (middle, Fig. \ref{lifting}a), a small bump appears 
within the valley, which at even smaller distance (bottom) increases in height and diameter. With respect to the neighbouring hill, the valley is 
lifted by 32\,pm, i.e. about half of the initial height difference hill-valley. Within the lifted area, the atomic structure changes from 
hexagons to bumps appearing at every second atom position, which has been checked by following up lines of atomic corrugation.
The symmetry between A and B lattice is broken, most probably due to in-plane compressive stress.
An explanation is sketched in Fig. \ref{lifting}c: the electrostatic and vdW force of the tip ($F_{\rm el}$, $F_{\rm vdW(tg)}$) lift the graphene valley until compensated by the restoring elastic force $F_{\rm mem}$ of the membrane and the vdW force $F_{\rm vdW(sg)}$ of the substrate.
Since the two tip forces change with lateral tip position, a dynamic image of the lifting results as indicated by white dots in Fig. \ref{lifting}c.
While a hill appears within the STM image, the membrane still maintains its valley-like shape, but with reduced curvature.
The resulting compressive, lateral force within the lifted area can be reduced by a vertical zig-zag atomic arrangement straightforwardly explaining the observed symmetry breaking between A and B lattice.

While the membrane's lifting in Fig. \ref{lifting} is reversible, other valleys exhibit hysteresis. Figure \ref{hysteresis}b shows the $I(z)$-curve of a hysteretic area at a sample voltage of $U=1$\,V.
While approaching or retracting the tip, a jump in current $I$ by three orders of magnitude is observed with a hysteresis of $50$\,pm. Such 
hysteretic membranes can be identified directly within constant-current images, if $I$ is chosen within the bistable range. 
This is demonstrated in Fig. \ref{hysteresis}a, where a valley (hill) showing atomic hexagons is visible at low (high) $I$, while a noisy area appears in between.
The noisy behaviour is caused by the feedback loop, i.e.\ the tip approaches which increases the tip forces and induces a snapping of the membrane towards the tip; this increases $I$ by three orders of magnitude and the tip retracts again leading to reduced tip forces and a snapping back of the membrane.
We observed bistability only in valleys and never on hills.
Although analysing several tens of valleys, we could not find any correlation between hysteretic or reversible behaviour and depth, width or curvature of the valleys.
Thus, we conclude that graphene-substrate interactions not visible by STM are crucial.

\begin{figure*}[htb]
\includegraphics[width=18cm]{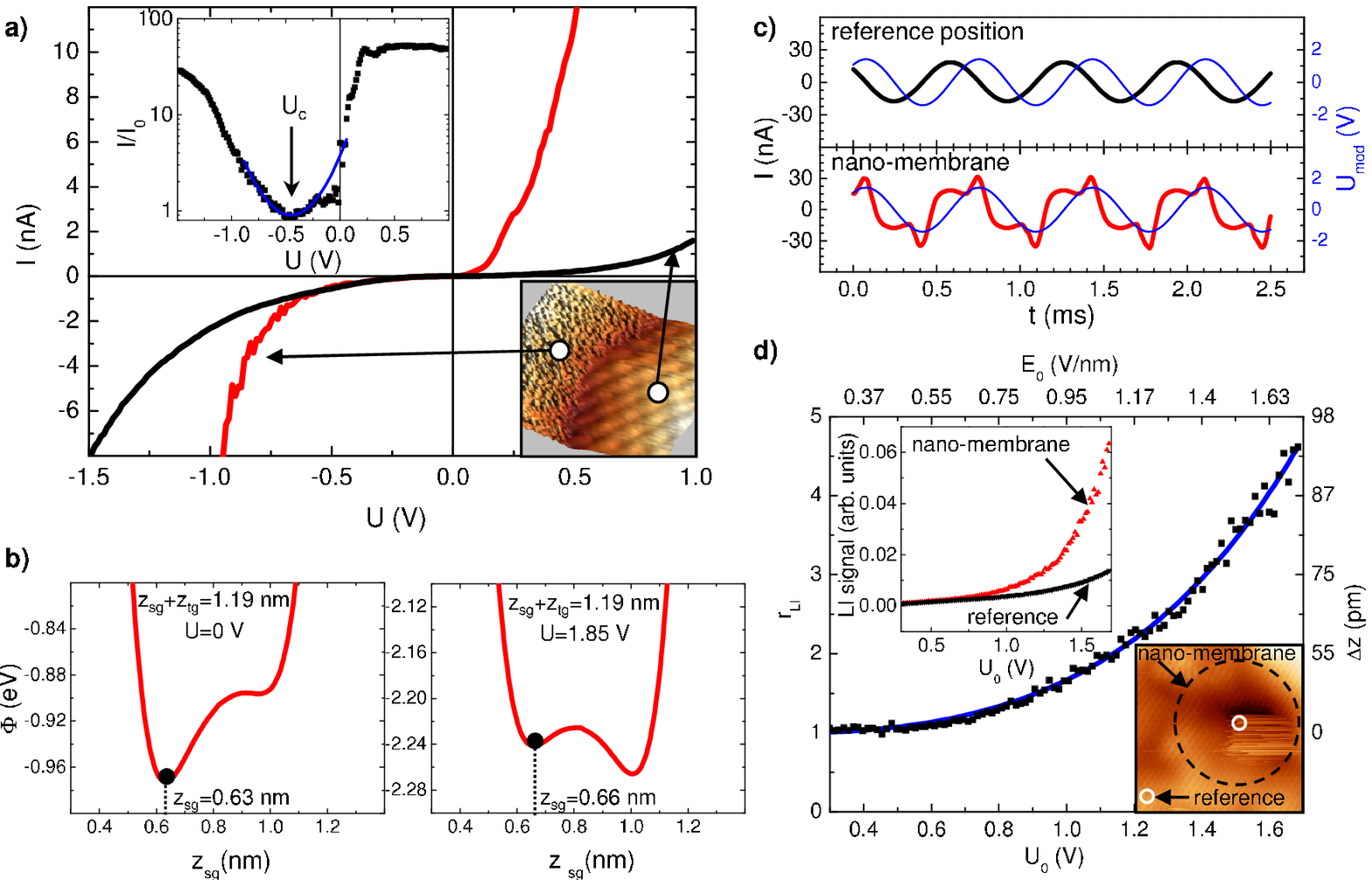}
\caption{\label{measurements}Oscillatory deflection of the graphene nano-membrane. a) $I(U)$-curve ($U_{\rm stab}=-0.3$\,V, $I_{\rm stab}=100$\,pA) 
recorded on a valley (red) and on a hill position (black) as marked in the lower inset. Upper Inset: logarithmic plot of the ratio of the two 
$I(U)$ curves being proportional to the deflection $ z_+(U)$. The minimum marks the contact potential $U_{\rm C}$; blue line: parabolic fit. b) Calculated 
potential $\Phi_{\rm sum}$ for an initial substrate-graphene distance $z_{\rm sg}=0.63$\,nm and tip-graphene distance
$z_{\rm tg}=0.56$ nm. Even at $U=1.85$\,V the system cannot move to the 
right minimum, but exhibits only a reversible shift of 30\,pm. c) Current response (black: hill, red: valley) to an oscillating tip voltage 
$U_{\rm mod}$ (blue) measured at reference (hill) and on nano-membrane (valley). d) Upper inset: in-phase lock-in signal recorded on reference 
(black) and nano-membrane (red). Main: ratio of the two lock-in signals $r_{\rm LI}$ from the upper inset (symbols) in comparison with a fit 
using the model of a prestrained clamped membrane (blue line) {\protect \cite{Lee}}. Right and upper scale show the deduced deflection amplitude 
$\Delta z$ (right axis) and the applied electric field amplitude $E_0$ (upper axis). Lower inset: STM image with the two measurement positions (white circles) and area of the membrane (dashed circle) 
marked.}
\end{figure*}

In order to model the observed behaviour, we analyse the involved interaction potentials.
Besides the electrostatic potential $\Phi_{\rm el}$ between tip and graphene, the vdW potentials $\Phi_{\rm vdW}$ for tip/graphene and  graphene/SiO$_2$ \cite{Israel}, and the elastic potential of the membrane $\Phi_{\rm mem}$ are considered (see methods).
They are plotted in Fig. \ref{hysteresis}c as a function of vertical graphene position for a fixed tip-substrate distance (1.63\,nm).
The summed up potential $\Phi_{\rm sum}$ exhibits two local minima representing the observed bistability.
Fig. \ref{hysteresis}d shows two nearly degenerate minima at $U=0$\,V and tip-substrate distance $1.46$\,nm (top).
The lower minimum transforms into a saddle at $U=1.3$\,V (middle) forcing the membrane to the upper minimum, i.e. 0.34 nm closer to the tip.
Increasing now the tip-graphene distance $z_{\rm tg}$ by 0.34\,nm (reduced $I$) shakes the potential back and the membrane flips back to its original minimum (bottom).
Note that $z_{\rm tg}\simeq 0.56$ nm is estimated by extrapolation of $I(z_{\rm tg})$ towards the contact conduction $G_0=2e^2/h$ \cite{Berndt} leaving only the initial substrate-graphene distance $z_{\rm sg}=0.9$ nm as a fit parameter.
By decreasing the initial $z_{\rm sg}$ to $0.63$\,nm, we still calculate two minima, but we are not able to switch the membrane to the upper position by reasonable $U$.
This explains the reversible behaviour observed in Fig. \ref{lifting}.

To explore the electric response of the membrane, we varied $U$.
First, we measured $I(U)$ on a hysteretic valley in comparison with $I_0(U)$ on a hill (Fig. \ref{measurements}a).
Assuming the same electronic structure at both positions, the ratio of the two curves is given by $I(U)/I_0(U)=\exp(-2\kappa(U)\cdot z_+(U))$, where $\kappa(U)$ is the electron's decay constant determined within the supplement.
Thus, the logarithmic plot of $I(U)/I(U_0)$ directly displays the deflection $z_+$ of the valley with respect to the hill  (upper inset, Fig. \ref{measurements}a).
The minimum of $z_+ (U)$ marks the absence of electrostatic force at the contact potential $U_{\rm C}=-0.45$\,V implying a tip work function of 5.11\,eV (graphene: 4.66\,eV \cite{Bin}).

Next, we apply an ac-voltage $U(t)=U_{\rm dc}+U_0\cos(2\pi\nu t)$ with varying amplitude $U_0$ at a dc-offset $U_{\rm dc}=-0.3$\,V $\simeq U_{\rm C}$ and fixed tip-substrate distance. 
The tip is placed above the nano-membrane marked by the dashed circle in the inset of Fig. \ref{measurements}d. We choose $I_{\rm stab}= 0.2$\,nA low enough to avoid snapping of the membrane. Figure \ref{measurements}c shows the resulting $I(t)$ and the applied $U_{\rm mod}(t)=U(t)-U_{\rm dc}$ for the membrane (bottom) and on a reference (hill) 
position (top). At reference, $I(t)$ is dominated by the capacitive crosstalk (phase shift of $90^\circ$). At the membrane, an additional in-phase 
signal, non-linear with respect to $U_{\rm mod}(t)$, indicates the reversible membrane movement. Fig. \ref{measurements}b displays the 
corresponding $\Phi_{\rm sum}(z_{\rm sg})$ at the two extrema of electrostatic force ($z_{\rm tg}=0.56$\,nm). The membrane does not 
switch between the two minima, but oscillates reversibly within one valley in agreement with experiment.

Finally, we use lock-in technique to measure the deflection amplitude $\Delta z$ with respect to $U_0$. The inset of Fig. \ref{measurements}d shows the lock-in output measured on the nano-membrane (red) and on a reference position (black).
The stronger signal at the membrane again indicates its oscillatory motion.
To get rid of unknown parameters we use the ratio of the two lock-in signals
\begin{eqnarray}\label{lockin}
r_{\rm LI}=\frac{\int_t^{t+T}I(t)\cos(\omega t)\,dt}{\int_t^{t+T}I_0(t)\cos(\omega t)\,dt}\, .
\end{eqnarray}
From $r_{\rm LI}$ we deduce $\Delta z$ numerically (see supplement) as displayed on the right scale of Fig. \ref{measurements}d.
The maximum amplitude is higher than expected from the model in Fig. \ref{measurements}b.
This is likely caused by the fact that also the hills are simultaneously lifted by the tip forces, which continuously decreases $\Phi_{\rm vdW(sg)}$ with respect to our model.
Nevertheless, the displayed $\Delta z$ and the corresponding electric field amplitude $E_0$ deduced from $U_0$ and $\Delta z$ (see supplement) give an impression of the amplitude-field relation achievable by external excitation.

The basic resonance frequency of the nano-membrane $\nu_0$ can be roughly estimated by clamped membrane theory using $E^{\rm 2D}=340$\,N/m \cite{Lee}, membrane radius $r=2.58$\,nm, and mass $m_0=1.59\times 10^{-23}$\,kg and assuming HOPG values for thickness $h=0.335$\,nm, and Poisson-ratio $s=0.16$ \cite{Robinson}:
\begin{eqnarray}\label{resonance}
\nu_0=\frac{h}{4r}\sqrt{\frac{\pi^3 E^{\rm 2D}}{3m_0(1-s^2)}}\simeq 430 {\rm GHz}\, .
\end{eqnarray}

More adequate molecular dynamics simulations (MDS) for monolayer graphene reveal $\nu_0 =400$\,GHz for a smaller area (4.3\,nm$^2$) exhibiting 75 \% deviation from eq. \ref{resonance} \cite{Inui}.
Anyhow, the nano-membranes consisting of only 200-800 atoms are ideal resonators for ultra sensitive mass detection. Adsorbing one hydrogen atom of mass $m_{\rm a}$ would lead to a relative frequency-shift of
\begin{eqnarray}\label{shift}
\frac{\Delta\nu}{\nu_0}=\left[\left(1+\frac{m_{\rm a}}{m_0}\right)^{-0.5}-1\right]\simeq 10^{-4}\, .
\end{eqnarray}

MDS finds quality factors $Q =2500$ $(273.000)$ for unsupported graphene monolayers at 300 (3)\,K \cite{Kim2}. Experiments reveal lower values 
for multilayer graphene probably due to interlayer friction \cite{Bunch,Sanchez, Bunch2, Peng}, but $Q=4000$ has been achieved  using oxidized 
multilayer graphene at 300\,K \cite{Robinson}. This might be sufficient to detect frequency shifts of $10^{-4}$. Measuring membrane oscillations 
would be eased by the nonlinearity of $I(z)$ requiring only dc detection, but the excitation in the range of 100\,GHz remains a technical 
challenge.

\section{methods}
The preparation of the graphene sample is done by mechanical exfoliation on a SiO${_2}$ substrate as described elsewhere \cite{Novo,Lemme}. A 
graphene flake containing a monolayer region is identified by an optical microscope. In addition, the film thickness is confirmed by Raman 
spectroscopy \cite{Geringer}. A gold contact surrounding the graphene is produced by e-beam lithography. In order to remove the residual resist 
and adsorbates as water, the sample is heated to 170$^\circ$C for a few hours, first in air, and then, directly before the measurement, in 
ultra-high vacuum ($p=2\times 10^{-10}$\,mbar). The monolayer region of $(18\times 26)\,\mu$m$^2$ is positioned below the tip of the STM by a 
piezo motor using the control by an optical long distance microscope with a resolution of 5\,$\mu$m. The measurements are performed with a high 
resolution STM operating at $T=4.8$\,K \cite{Mashoff}. The tip is prepared prior to measurements by applying voltage pulses and field emission on 
the gold contact region. All STM images are measured in constant-current mode with the voltage $U$ applied to the sample. All spectroscopic 
curves are measured with the feedback switched off at stabilization voltage $U_{\rm stab}$ and stabilization current $I_{\rm stab}$. Afterwards, 
either $U$ or the tip sample distance $z$ is changed, while measuring the tunnelling current $I$.

Calculations of the interaction potentials acting on the graphene membrane are described in detail in the supplement. In short, we use a 
dielectric plane for SiO$_2$, a metallic circle of cosine shape in vertical direction for the membrane and a metallic W tip of parabolic shape 
(central radius: $0.7$\,nm) to calculate the vdW potentials \cite{Israel}.
The elastic potential $\Phi_{\rm mem}$ is modelled by a clamped membrane implying $\Phi_{\rm mem} = A \cdot z_+^4$  with $A$ taken from 
experiment \cite{Lee} and $z_+$ being the membrane deflection. We assume two minima (two relaxed membrane positions) of two $z_+^4$-curves with origins separated by 0.12\,nm in 
order to describe the fact that the membrane is first laterally compressed by lifting with respect to the hills, but is relaxed again after being 
higher than the surrounding hills. We do not include a pretension term \cite{Lee}, since pretension is given explicitly by the vdW forces. The electrostatic 
energy $\Phi_{\rm el}=1/2\cdot C_{\rm m} U_{\rm eff}^2$ is calculated using $C_{\rm m}$ as the tip-nano-membrane capacitance modelled as a sphere 
in front of a plane and $U_{\rm eff}$ as the voltage drop between tip and graphene reduced with respect to $U$ due to graphene's finite carrier 
density.

\section{acknowledgement}
We appreciate financial support of the German Science foundation via Mo 858/8-1.

\section{Supplementary information}
\subsection{Determination of the initial tip/graphene distance $z^0_{\rm tg}$ and the decay constant $\kappa(U)$}

The absolute distance between tip and graphene obtained after stabilization, $z^0_{\rm tg}$, could be determined, in principle, by decreasing the distance 
between tip and sample until the conductance reaches the conductance quantum $G_0=2e^2/h$ \cite{Berndt}. Unfortunately, this does not work for graphene, because the graphene is lifted during the tip approach, even on the hills (reference positions) as indicated from significantly too large decay constants 
$\kappa$ extracted from $I(z)$-spectra, which have been measured systematically on the graphene surface. The only way to reasonably estimate $z^0_{\rm tg}$ 
is to calculate $\kappa=23.3$\,nm$^{-1}$ with the help of equation \ref{kappa} (below) and, then,
to extra\-po\-late the exponential behaviour of the conductance with respect to the distance by using the stabilization 
parameters $U_{\rm s}=-0.6$\,V and $I_{\rm s}=200$\,pA:
\begin{eqnarray}
\frac{I_{\rm s}}{U_{\rm s}}&=&\frac{2e^2}{h}\exp\left\lbrace -2\kappa (U_{\rm s}) z^0_{\rm tg}\right\rbrace \nonumber\\
\Leftrightarrow z^0_{\rm tg}&=&-\frac{1}{2\kappa (U_{\rm s})}\ln\left\lbrace\frac{1}{G_0}\frac{I_{\rm s}}{U_{\rm s}}\right\rbrace=0.56 \hbox{\,nm}.
\end{eqnarray}

To determine the decay constant $\kappa(U)$, we use a planar tunnelling junction with a correction factor $\xi$ \cite{Ukraintsev}:
\begin{eqnarray}\label{kappa}
\kappa(U)=\xi\sqrt{\frac{2m_{\rm e}}{\hbar^2}\left(\phi_{\rm eff}-\left|\frac{eU}{2}\right|\right)}.
\end{eqnarray}
The effective work function $\phi_{\rm eff}=(\phi_{\rm gr}+\phi_t)/2=4.89$\,eV has been calculated using the known graphene work function of $\phi_{\rm 
gr}=4.66$\,eV \cite{Bin} and a tip work function of $\phi_t=5.11$\,eV derived from the measured contact potential as displayed in Fig. 3a of the main text. The correction factor 
$\xi=1.06$ has been determined by fitting $I(z)$-curves measured with the same microtip on Au(111)  by equation \ref{kappa} (work function: $\phi_{\rm Au}=5.31$\,eV 
\cite{Mich}).

\subsection{Calculation of the interaction potentials acting on the nano-membrane}

In order to describe the observed behaviour of the nano-membrane, the involved interaction potentials are analyzed in detail. Besides the 
electrostatic potential $\Phi_{\rm el}$ induced by the tip, the Casimir/van-der-Waals potentials $\Phi_{\rm vdW}$ induced by the tip and the 
SiO$_2$-substrate as well as the elastic restoring force of the membrane itself $\Phi_{\rm mem}$ are considered.

\subsubsection{Casimir/van-der-Waals potential induced by the tip}

The description of the van-der-Waals and Casimir potential $\Phi_{\rm vdW}$ per unit area $A$ between two materials is given by \cite{Bordag}:
\begin{eqnarray}\label{casimir}
\lefteqn{\frac{\Phi_{\rm vdW}}{A}(z_{\rm tg})=}\nonumber\\
& & \frac{\hbar}{4\pi^2}\int_0^\infty k_{\perp}dk_{\perp}\int_0^\infty\ln\left[1-r_{\rm gr}r_{\rm w}e^{-2qz_{\rm tg}}\right]d\omega
\end{eqnarray}
where $r_{\rm gr}$ and $r_{\rm w}$ denote the frequency dependent reflection coefficients of graphene and tungsten, respectively, $k_{\perp}$ is the wave number parallel to the 
surface, $z_{\rm tg}$ is the variable distance between graphene and the tip apex and $\omega$ is the frequency ($\hbar$: Planck's constant). The reflection coefficient of graphene can be calculated by \cite{Bordag}:
\begin{eqnarray}
r_{\rm gr}=\frac{c^2q\Omega}{c^2q\Omega+\omega^2}\, ,
\end{eqnarray}
with $\Omega=6.75\times 10^5$m$^{-1}$, $c= 3\cdot 10^8$ m/s and
\begin{eqnarray}
q=\sqrt{k_{\perp}^2+\frac{\omega^2}{c^2}}\, .
\end{eqnarray}

To determine the reflection coefficient of the tungsten tip, the frequency dependent dielectric constant $\varepsilon(\omega)$ has to be used:
\begin{eqnarray}\label{reflection}
r_{\rm w}=\frac{\varepsilon(\omega)q-k}{\varepsilon(\omega)q+k}
\end{eqnarray}
with
\begin{eqnarray}\label{kvalue}
k=\sqrt{k_{\perp}^2+\varepsilon(\omega)\frac{\omega^2}{c^2}}\, .
\end{eqnarray}

The dielectric function can be approximated knowing the plasma frequency of tungsten $\omega_{\rm p,W}=9.74\times 10^{15}$\,Hz \cite{Ordal}: 
\begin{eqnarray}
\varepsilon(\omega)=1+\frac{\omega_{\rm p,W}^2}{\omega^2}.
\end{eqnarray}
The total interaction potential $\Phi_{\rm vdW}(z)$ is determined by an integration of equation \ref{casimir} over the circular area of the nano-membrane:
\begin{eqnarray}\label{iapotential}
\Phi_{\rm vdW}(z_{\rm tg}(r'=0))=2\pi\int^{r}_0\frac{\Phi_{\rm vdW}}{A}(z_{\rm tg}(r'))r'dr',
\end{eqnarray}
where $z_{\rm tg}(r')$ denotes the vertical distance between tip and graphene at a lateral distance $r'$ measured from the centre of the membrane and $z_{\rm tg}(r'=0)$ indicates that we use the tip-graphene distance in the centre of the membrane
as the variable for $\Phi_{\rm vdW}$. Assuming a circular membrane with the measured radius of $r=2.58$\,nm and a 2D-cosine shaped corrugation (see figure S1), as well as a parabolic tip with central radius $R$, $z_{\rm tg}(r')$ can be described as:
\begin{eqnarray}
z_{\rm tg}(r')=\frac{r'^2}{2R}+\left[z_{\rm tg}(0)-z^{\rm hill}_{\rm tg}\right]\cos^2\left(\frac{\pi r'}{2r}\right)+z^{\rm hill}_{\rm tg} \, .
\end{eqnarray}

The vertical distance between tip apex and the hills surrounding the membrane $z^{\rm hill}_{\rm tg}$ is determined by analyzing the measured valley depth with respect to the surrounding hills.
For the tip radius $R$ we can only give an upper limit of $R=2.3$\,nm determined by analyzing STM images of atomically resolved valleys, which would not have been resolved by larger tips due to convolution effects.
The smallest possible tip radius of $R=0.3$\,nm is given by a tetraedric alignment of the first four atoms.
For the calculation, we use a value of $R=0.7$ nm, determined as described below.

\begin{figure}\label{distances}
\includegraphics[width=8.5cm]{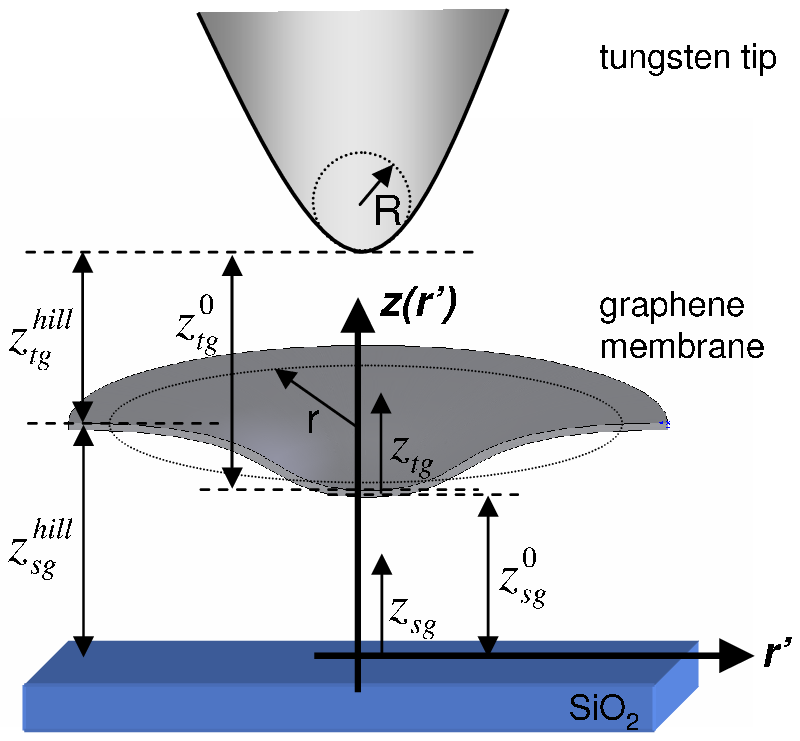}
{S1: Definition of distances used in the calculation of the interaction potentials.}
\end{figure}

\subsubsection{Casimir/van-der-Waals potential induced by the SiO$_2$-substrate}

The interaction potential between graphene and the amorphous SiO$_2$-subtrate $\Phi_{\rm vdW(sg)}$ is calculated similarly 
using equation \ref{casimir}$-$\ref{kvalue} and \ref{iapotential}, but replacing $z_{\rm tg}(r')$ by the distance between graphene and the SiO$_2$ substrate $z_{\rm sg}(r')$ as well as $r_{\rm w}$ by the reflection coefficient of SiO$_2$ $r_{\rm SiO2}$. In case of an insulating material like SiO$_2$, $\varepsilon(\omega)$ is given by another expression \cite{Bordag}:
\begin{eqnarray}
\varepsilon(\omega)=1+\frac{\varepsilon(0)-1}{1+\frac{\omega^2}{\omega_{\rm e}^2}}\, ,
\end{eqnarray}
where $\omega_{\rm e}=1.05\times 10^{16}\, {\rm s}^{-1}$ is the main electronic absorption frequency being within the ultra violet region \cite{Israel,Heraeus}. For amorphous SiO$_2$, we use the known dielectric constant at zero frequency of $\varepsilon(0)=3.9$ \cite{Wilk}.
To describe $z_{\rm sg}(r')$, assuming a plane substrate and again a 2D-cosine shaped membrane as sketched in Fig.\ S1, we use
\begin{eqnarray}
z_{\rm sg}(r')=z^{\rm hill}_{\rm sg}-\left[z^{\rm hill}_{\rm sg}-z_{\rm sg}(0)\right]\cos^2\left(\frac{\pi r'}{2r}\right)
\end{eqnarray}
with $z^{\rm hill}_{\rm sg}$ being the distance between substrate and the borders of the membrane (surrounding hills) and $z_{\rm sg}(0)$ being the distance between the substrate and the lowest point of the membrane.
The value of $z_{\rm sg}(0)$ can vary with the applied voltage $U$ or with the tip-substrate distance.
However, the change of $z_{\rm sg}(0)$ with $U$ or tip-substrate distance can be measured via the tunnelling current.
Only the offset of the initial tip-substrate distance, $z_{\rm sg}^0$, without electric field has to be taken as a fit parameter.

\subsubsection{Elastic membrane potential}

In order to determine the elastic potential $\Phi_{\rm mem}$ of the nano-membrane, we assume a cubic force dependence of the deflection $z_+$ as given by the classical clamped membrane theory according to Komaragiri et al. \cite{Komaragiri}\cite{Lee}. 
The nanomembrane gets laterally compressed if lifted until the centre of the membrane is at the same height as the surrounding hills (see Fig. 1c of the main text). If lifted further the membrane gets relaxed again up to a second stable position above the surrounding hills. In order to model this behaviour, we compose the membrane potential $\Phi_{\rm mem}$ of two parts $\Phi^{+}(z_{\rm sg}(0))$ and $\Phi^{-}(z_{\rm sg}(0))$ with minima vertically symmetric with respect to the position of largest compression leading to:
\begin{eqnarray*}
\Phi^+(z_{\rm sg})&=&-0.265E^{\rm 2D}\frac{(z_{\rm sg}-z^{\rm hill}_{\rm sg}+\frac{d}{2})^4}{r^2}, \hbox{if~} z_{\rm sg}<z^{\rm hill}_{\rm sg},\\
\Phi^-(z_{\rm sg})&=&-0.265E^{\rm 2D}\frac{(z_{\rm sg}-z^{\rm hill}_{\rm sg}-\frac{d}{2})^4}{r^2}, \hbox{if~} z_{\rm sg}>z^{\rm hill}_{\rm sg},
\end{eqnarray*}
where $E^{\rm 2D}$ denotes the two dimensional Young's modulus due to compressing or stretching of the atomic bonds, which has been measured previously to be $E^{\rm 2D}=340$\,N/m \cite{Lee}.
The distance between the two potential minima $d$ is the second free parameter of our model, only limited to about twice the valley depth. We found that $d=0.12$ nm is able to reproduce the behaviour of the membranes displayed in Fig. 1 and 2 of the main text. This number is lower than the valley depth appearing in the STM images implying that the van-der-Waals force of the SiO$_2$ substrate increases the corrugation strength within the graphene flake.

\subsubsection{Electrostatic potential induced by the tip}

The electrostatic potential $\Phi_{\rm el}$ induced by the tip is given by
\begin{eqnarray}
\Phi_{\rm el}=\frac{1}{2}C_{\rm m} U_{\rm eff}^2,
\end{eqnarray}
with the capacitance $C_{\rm m}$ and the effective gap voltage $U_{\rm eff}$ to be determined. We approximate the tip-sample system by a capacitor consisting of a sphere with radius $R$ above a circular plate of radius $r$ corresponding to tip and membrane, respectively.
The capacitance $C$ for $r=\infty$ can be calculated analytically resulting in \cite{Smythe}:
\begin{eqnarray}\label{capacity}
\lefteqn{C=4\pi \varepsilon\varepsilon_0 R\sinh\left[R\cosh\left(\frac{z_{\rm tg}+R}{R}\right)\right]}\nonumber\\
 & & \cdot\sum_{n=1}^{\infty} \sinh^{-1}\left[nR\cosh\left(\frac{z_{\rm tg}+R}{R}\right)\right],
\end{eqnarray}
with the distance between the tip apex and the infinite plate $z_{\rm tg}= z_{\rm tg}(0)$. The capacitance determined by equation \ref{capacity} has to be modified because of the finite area of the nano-membrane below the tip. Therefore, we assume a Gaussian shape of the total charge density of the infinite plate with a maximum at $r'=0$ to be determined below and calculate the capacitance of the finite membrane by integration only over the circular area of the nano-membrane. We end up with a capacitance $C_{\rm m}$ of
\begin{eqnarray}\label{capacitymod}
C_{\rm m}=C\left[1-\exp\left(-\frac{\pi\varepsilon\varepsilon_0 r^2}{Cz_{\rm tg}}\right)\right]\, .
\end{eqnarray}

Because of the finite charge carrier concentration of graphene, the effective voltage $U_{\rm eff}$ between the tip and graphene is smaller than the applied bias-voltage $U$.
The remaining voltage leads to a considerable Fermi level shift within the graphene until the charge carrier density is high enough to screen the electric field.  
As a consequence, there is a potential drop between the graphene just below the STM tip and the gold electrode connected to the external power supply.
Due to the linear dispersion relation of graphene the two dimensional charge carrier density $n$ can be written as \cite{Gusy}:
\begin{eqnarray}\label{n}
n=\frac{e^2(\tilde{U}-U_{\rm eff})^2}{\pi\hbar^2 v_{\rm F}^2}\, ,
\end{eqnarray}
where $\tilde{U}:=U-U_{\rm C}$ with $U_{\rm C}$ being the contact potential determined in Fig.\ 3a of the main text,
$v_{\rm F}=1.1\times 10^6$\,m/s is the Fermi-velocity of graphene and $e$ is the electron's charge. In equilibrium, the electrons screen the electric field $E$ and the resulting 2D charge density can be most easily approximated by a plate capacitor leading to:
\begin{eqnarray}\label{rho}
ne=\varepsilon\varepsilon_0 E(U_{\rm eff})=\varepsilon\varepsilon_0 \frac{U_{\rm eff}}{z^+_{\rm tg}}.
\end{eqnarray}
Thereby, $z^+_{\rm tg}$ denotes the distance between the plate and the centre of mass of the lower half sphere approximating the tip, which has been chosen to map the model of two parallel plates to the model of a sphere and a plate. With the help of equations \ref{n} and \ref{rho}, the effective voltage drop between tip and graphene becomes:
\begin{eqnarray}\label{Ueff}
U_{\rm eff} &=& \tilde{U} + \frac{\alpha}{2z^+_{\rm tg}} - \sqrt{\frac{\alpha^2}{4(z^+_{\rm tg})^2}+\frac{\alpha \tilde{U}}{z^+_{\rm tg}}}\, ,\\
\alpha &=& \frac{\varepsilon\varepsilon_0\pi\hbar^2 v_{\rm F}^2}{e^3}\, . \nonumber
\end{eqnarray}

The value of the dielectric constant for graphene on SiO$_2$ has been calculated previously using the image potential method and amounts to $\varepsilon=2.5$ \cite{Xu}. The resulting $n$ calculated straightforwardly by inserting $U_{\rm eff}$ into equation \ref{n} has been used self-consistently as the maximum of the Gaussian charge density required to calculate $C_{\rm m}$.

\subsection{Excitation of the nano-membrane by ac-voltage}

We applied an ac-voltage $U(t)=U_{\rm dc}+U_0\cos(2\pi\nu t)$ with a varying amplitude $U_0$, at a dc-offset $U_{\rm dc}$ and a frequency of $\nu=1.4$\,kHz. In addition, we define the voltage $U^{\rm c}(t)=U_{\rm dc}+U_0\cos(2\pi\nu t)-U_{\rm c}$ including the contact potential $U_{\rm c}$ determined from Fig.\ 3a of the main text. The time dependent tunnelling currents $I_0(t)$, measured at the stable reference position and $I(t)$ measured above the nano-membrane can be described using the linear graphene density of states as:
\begin{eqnarray}\label{I0}
I_{0}(t)\propto \frac{U(t)^3}{|U(t)|} \exp\left\lbrace-2\kappa (U(t)) z^0_{\rm tg}\right\rbrace
\end{eqnarray}
and
\begin{eqnarray}\label{I}
I(t)\propto \frac{U(t)^3}{|U(t)|} \exp\left\lbrace-2\kappa (U(t))\left[z^0_{\rm tg}- z_+(U_{\rm eff}(t))\right]\right\rbrace,
\end{eqnarray}
where $z_+(U_{\rm eff})$ denotes the deflection of the nano-membrane with respect to its position in the absence of electric field and $U_{\rm eff}(t)$ is the part of the corrected voltage $U^{\rm c}(t)$ dropping between membrane and tip. Note, that the expression for $z_+$ is assumed not to be present on the reference position. 
The quadratic dependence of the tunnelling current with respect to the bias voltage is derived from the linear dispersion relation of graphene and has been checked by according fits to the measured $I(U)$-spectra.
In Fig.\ 3d of the main text, we display the lock-in ratio:
\begin{eqnarray}\label{rli}
r_{\rm LI}=\frac{\int_t^{t+T}I(t)\cos(2\pi\nu t)\,dt}{\int_t^{t+T}I_0(t)\cos(2\pi\nu t)\,dt},
\end{eqnarray}
which can be directly used to determine the deflection amplitude $\Delta z$ numerically as displayed on the right of Fig.\ 3d of the main text. Accordingly, the upper scale of Fig.\ 3d of the main text shows the electric field amplitude $E_0$ given by $E_0=U_{\rm eff,max}/z_{\rm tg,min}$ with $z_{\rm tg,min}$ being the minimal distance between graphene and tip and $U_{\rm eff,max}$being the maximal effective voltage during the oscillation.

Finally, we describe our estimate of the tip radius $R$. Therefore, we use the simplified clamped membrane model for the force-deflection curve, consisting of a linear term caused by so-called pretension and a cubic term describing the compression of the atomic bonds given by $E^{\rm 2D}$ \cite{Lee}. The equilibrium between the electrostatic force and the elastic membrane force is then given by:
\begin{eqnarray}\label{force}
\frac{1}{2}\frac{\partial C_{\rm m}}{\partial z_{\rm tg}}(U_{\rm eff}(t))^2=-\sigma_0^{\rm 2D}\pi z_+-E^{\rm 2D}1.06\frac{z_+^3}{r^2},
\end{eqnarray}
where $\sigma_0^{\rm 2D}$ denotes the two dimensional pretension. After solving towards $z_+ (U_{\rm eff})$ numerically, we fitted the measured $r_{\rm LI}$ taking $\sigma^{\rm 2D}_0$ and the tip radius $R$ as the only free parameters. The excellent fit displayed in Fig. 3d of the main text results in a tip radius of $R=0.7$\,nm and a pretension of $\sigma^{\rm 2D}_0=0.62$\,N/m.

\end{document}